\documentclass[11pt,amssymb,nofootinbib]{revtex4}
\usepackage[latin9]{inputenc}
\usepackage{textcomp}
\usepackage{amsmath}
\usepackage{amssymb}
\usepackage{esint}
\usepackage{graphicx}
\usepackage{epstopdf}
\usepackage[english]{babel}

\makeatletter
\@ifundefined{textcolor}{}
{%
 \definecolor{BLACK}{gray}{0}
 \definecolor{WHITE}{gray}{1}
 \definecolor{RED}{rgb}{1,0,0}
 \definecolor{GREEN}{rgb}{0,1,0}
 \definecolor{BLUE}{rgb}{0,0,1}
 \definecolor{CYAN}{cmyk}{1,0,0,0}
 \definecolor{MAGENTA}{cmyk}{0,1,0,0}
 \definecolor{YELLOW}{cmyk}{0,0,1,0}
}

\usepackage{amsthm}\usepackage{amscd}\usepackage{bm}\usepackage{bbm}

\let\vec\boldvec%

\makeatother

\begin{document}

\title{On the spontaneous emission of electromagnetic radiation in the CSL model}


\author{Sandro Donadi}

\email{sandro.donadi@ts.infn.it}

\affiliation{Department of Physics, University of Trieste, Strada Costiera 11,
34151 Trieste, Italy}

\affiliation{Istituto Nazionale di Fisica Nucleare, Trieste Section, Via Valerio
2, 34127 Trieste, Italy}

\author{Dirk-Andr\'e Deckert}

\email{deckert@math.ucdavis.edu}

\affiliation{Department of Mathematics, University of California, One Shields Ave, 95616 Davis, USA}

\author{Angelo Bassi}

\email{bassi@ts.infn.it}

\affiliation{Department of Physics, University of Trieste, Strada Costiera 11,
34151 Trieste, Italy}

\affiliation{Istituto Nazionale di Fisica Nucleare, Trieste Section, Via Valerio
2, 34127 Trieste, Italy}


\begin{abstract}
Spontaneous photon emission in the Continuous Spontaneous Localization (CSL) model is studied one more time. In the CSL model each particle interacts with a noise field that induces the collapse of its wave function. As a consequence of this interaction, when the particle is electrically charged, it radiates. As discussed in~\cite{ref:abd}, the formula for the emission rate, to first perturbative order, contains two terms: One is proportional to the Fourier component of the noise field at the same frequency as that of the emitted photon and one is proportional to the zero Fourier component of the noise field. As discussed in previous works, this second term seems unphysical. In~\cite{ref:abd}, it was shown that the unphysical term disappears when the noises is confined to a bounded region and the final particle's state is a wave packet. Here we investigate the origin of the unphysical term and why it vanishes according to the previous prescription. For this purpose, the electrodynamic part of the equation of motion is solved exactly while the part due to the noise is treated perturbatively. We show that the unphysical term is connected to exponentially decaying function of time which dies out in the large time limit, however, approximates to 1 in the first perturbative order in the electromagnetic field. \end{abstract}
\maketitle

\section{Introduction}

Models of spontaneous wave function collapse~\cite{ref:grw,ref:precsl,ref:csl,ref:rep1,ref:rep2,ref:Pearle1,ref:Pearle2,ref:qmupl,ref:qmuplfp} explicitly modify the Schr\"odinger equation.
New terms are added to the usual quantum Hamiltonian which describe the nonlinear interaction between any system and a classical noise field. The effect of the new terms is to collapse the wave function in space. They are engineered in such way to give only small deviations from the Schr{\"o}dinger dynamics for microscopic systems. However, their effect grows with the size of the system such that for macroscopic objects the collapse of the wave function is dominant. In this way the possibility of having superpositions of different macroscopic states, like those described in the Schr{\"o}dinger's cat gedanken experiment, is avoided, and the quantum measurement problem vanishes.

The deviation from the standard Schr{\"o}dinger dynamics implies that collapse models make  predictions which differ from quantum mechanical ones. Therefore, they can be tested experimentally~\cite{ref:rep2,ref:sci,ref:ap}. So far the process of spontaneous emission of electromagnetic radiation from charged particles sets the strongest upper bound on the collapse parameters~\cite{ref:ap,ref:sci} among all possible tests. The idea is very simple. Charged particles are always subject to a random motion due to the interaction with the noise causing the collapse. Therefore they emit radiation even when according to standard quantum mechanics there should not be any.

This problem has been considered several times in the literature~\cite{ref:fu,ref:ar,ref:bd,ref:abd} because of a discrepancy in the theoretical formulas, which have been derived using different techniques which are summarized in the introductory section of~\cite{ref:abd}.
In a nutshell, the situation is the following. When the emission rate  is computed by using  standard perturbation theory, two terms are present at the first perturbative order: The first one, as expected, is proportional to the Fourier component of the correlation function of the noise field computed at the same frequency as that of the emitted photon. The second term is instead proportional to the zero Fourier component of the noise field's correlation. This latter term is unexpected and seems unphysical. 

The analysis carried out in~\cite{ref:abd} shows that a way to avoid such an unphysical term is to use wave packets instead of plane waves and to confine the noise field to a bounded region. This means that standard perturbation theory cannot be naively applied to collapse models but leaves open the question of the precise (mathematical) origin of the unphysical term.

In~\cite{ref:ad} the same problem was considered within the mathematically simpler QMUPL (Quantum Mechanics with Universal Position Localizations) model~\cite{ref:rep2,ref:qmupl,ref:qmuplfp}, which allows for an analytical treatment of the radiation process. It was shown that the undesired term is associated to an exponentially vanishing function of time. At first perturbative order it survives simply because the exponential is approximated by 1. This result is true if the particle is bounded -- no matter how weakly. 

In this paper we apply the same methodology used for the QMUPL model to the structurally different CSL model. We consider the CSL model of a charged particle bounded by a harmonic potential. 
Apart from the dipole approximation, we solve the equation of motion for the particle's electrodynamic part of the Hamiltonian {\it exactly} while we treat the interaction with the noise field perturbatively. We show that the unphysical term does not appear -- even when considering plane waves and without bounding the noise to a finite region. Our analysis implies that the unphysical term, which is present in the standard first perturbative order, is canceled when higher order terms are included in the perturbative series. It appears fictitiously in the first perturbative analysis. We will  comment on the picture which emerges from this analysis and on the relation between with the result of~\cite{ref:abd}.

\section{The CSL model and the problem of spontaneous photon emission from a free particle}

We briefly review the CSL model, introduce the problem of radiation emission and summarize the results discussed in the literature.

\subsection{The CSL model for charged particles}

For a given Hamiltonian $H$ the CSL modified Schr\"odinger equation~\cite{ref:precsl,ref:csl,ref:rep1,ref:rep2} reads: 
\begin{equation}
d|\psi_{t}\rangle=\left[-\frac{i}{\hbar}Hdt+\frac{\sqrt{\gamma}}{m_{0}}\int d\mathbf{x}\,[M(\mathbf{x})-\langle M(\mathbf{x})\rangle_{t}]dW_{t}(\mathbf{x})-\frac{\gamma}{2m_{0}^{2}}\int d\mathbf{x}\,[M(\mathbf{x})-\langle M(\mathbf{x})\rangle_{t}]^{2}dt\right]|\psi_{t}\rangle.\label{eq:csl-massa}
\end{equation}
Matter is treated non-relativistically, which for our analysis is completely justified.
The second and third term on the right hand side of Eq.~\eqref{eq:csl-massa} induce the collapse of the wave function in the position basis. They contain the parameter $\gamma$, a positive coupling constant which sets the strength of the collapse, and the reference mass $m_{0}$, which is taken to be equal to that of a nucleon. The operator $M({\bf x})$ is given by the mass density of the second quantized matter field.
In this article we however focus on the behavior of a single non-relativistic particle only, and hence, limit our analysis to the one-particle sector of the Fock space of the matter field. In consequence, the operator $M({\bf x})$ takes the form:
\begin{equation}
M\left(\mathbf{x}\right)=mg\left(\mathbf{x}-\hat{\mathbf{q}}\right)\;\;\;\;\;\;\;\textrm{where}\;\;\;\;\;\;\;g(\mathbf{x})\;=\;\frac{1}{\left(\sqrt{2\pi}r_{C}\right)^{3}}\; e^{-\mathbf{x}^{2}/2r_{C}^{2}}.\label{eq:nnbnm}
\end{equation}
Here, $m$ denotes the mass of the particle, $\hat{\mathbf{q}}$ the position operator, and $r_{C}$ the second new phenomenological constant of the model\footnote{Note that $M(\mathbf{x})$ does not contain photon operators since we are assuming that the spontaneous collapse process occurs only for massive particles. This is the standard choice for the CSL model.}. $W_{t}\left(\mathbf{x}\right)$ is an ensemble of independent Wiener processes, one for each point in space, which are responsible for the random character of the evolution; the quantum average $\langle M(\mathbf{x})\rangle_{t}=\langle\psi_{t}|M(\mathbf{x})|\psi_{t}\rangle$ is responsible for its nonlinear character.

The first term on the right hand side of Eq.~\eqref{eq:csl-massa} describes the unitary part of the evolution which is governed by the Hamiltonian $H$. As anticipated, we study a system of a single non-relativistic charge coupled to a harmonic potential and a second-quantized electrodynamic field so that 
\begin{equation}
H \; = \; \frac{1}{2m}({\bf p} - e {\bf A})^2 + \frac{1}{2}
m\omega_0^2\, {\bf q}^2 + \frac{1}{2}\epsilon_0 \int d^3x \, \left[ {\bf
E}_{\perp}^2 + c^2 {\bf B}^2 \right].\label{eq:H}
\end{equation}
In Eq.~\eqref{eq:H} the first term represents the dispersion relation of the particle minimally-coupled to the electromagnetic field, where ${\bf A}({\bf x},t)$ denotes the electromagnetic vector potential. The second term models an external harmonic potential with angular frequency $\omega_0$. This potential can later be removed by $\omega_0\to0$. Finally, the last term is the dispersion relation of the electromagnetic field (in Coulomb gauge), where $\mathbf{E}_{\perp}$ denotes the transverse part of the electric component and ${\bf B}$ the magnetic one.

\subsection{Standard perturbative approach for the spontaneous photon emission rate of a free particle and its problems}

The details of how to compute the photon emission rate to first perturbative order with the CSL model can be found in~\cite{ref:abd}. Here we however wish to recapitulate some important steps. In order to simplify the whole computation, the starting point is to note that the master equation associated to Eq.~\eqref{eq:csl-massa} can also be derived from a standard Schr{\"o}dinger equation with a random Hamiltonian: 
\begin{equation}
H_{\text{\tiny TOT}}=H-\frac{\hbar\sqrt{\gamma}}{m_{0}} \int M(\mathbf{x})\xi_{t}(\mathbf{x})\, d^{3}x,\, 
\end{equation}
where $\xi_{t}(\mathbf{x})=dW_{t}(\mathbf{x})/dt$ is a white noise field with correlation $\mathbb{E}[\xi_{t}(\mathbf{x})\xi_{s}(\mathbf{y})]=\delta(t-s)\delta({\bf x-y})$.
The evolution is hence linear, thought stochastic. In consequence, it does not lead to state vector reduction. Nevertheless, since it reproduces the same master equation as that associated to Eq.~\eqref{eq:csl-massa}, and since any physical quantity that can be computed via the density matrix,  it makes no practical difference whether to use Eq.~\eqref{eq:csl-massa} or the standard Schr\"odinger equation with the Hamiltonian  $H_{\text{\tiny TOT}}$ in order to computed the average value of observables. The second possibility however allows for the use of the tools of standard perturbation theory.

One then identifies the unperturbed Hamiltonian as that of the particle (interaction with the noise field excluded) plus the kinetic term
of the electromagnetic field: 
\begin{equation}
{H}_{0}\;=\;\frac{{\bf p}^2}{2m} + \frac{1}{2}
m\omega_0^2\, {\bf q}^2 + \frac{1}{2}\epsilon_0 \int d^3x \, \left[ {\bf
E}_{\perp}^2 + c^2 {\bf B}^2 \right],
\end{equation}
whose eigenstates and eigenvalues are known. The perturbed part of the Hamiltonian contains both the matter-radiation interaction and the noise interaction: 
\begin{equation}
{H}_{1}\;=\; -\frac{e}{m}\mathbf{A}\cdot\mathbf{p}+\frac{e^{2}}{2m}\mathbf{A}^{2}- \hbar \frac{\sqrt{\gamma}}{m_{0}} \int d^{3}x \, M({\bf x}) \xi_{t}(\mathbf{x}).\label{eq:dsgfdj}
\end{equation}
The Feynman rules  for this dynamics are reported in~\cite{ref:abd}. Since one carries out the computation to the lowest order for both interactions one can neglect the second term in Eq.~\eqref{eq:dsgfdj}. To this perturbative order, the six Feynman diagrams shown in Fig. 1 have to be considered.
\begin{figure}
{\includegraphics[width=16.5cm, keepaspectratio]{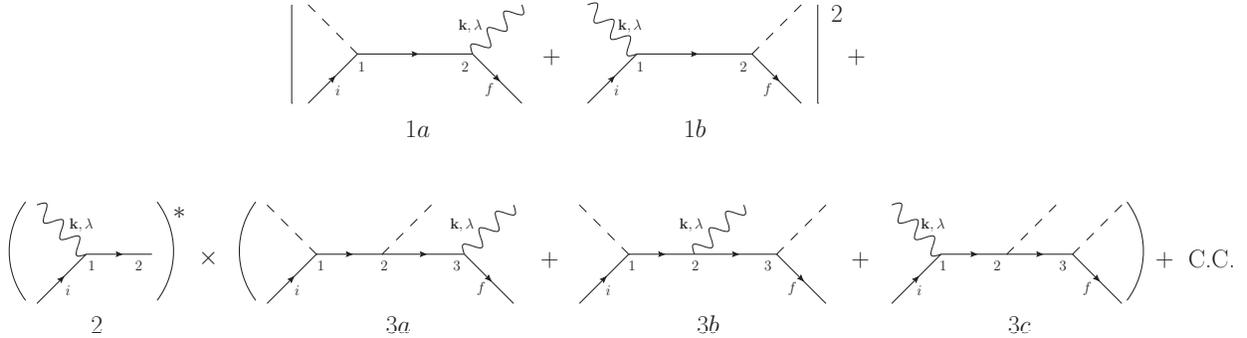}}
\caption{Lowest order contributions to the emission rate according to the CSL model. ``C.C." denotes the complex conjugate of the term in the second line. Solid lines represent the charged fermion, wavy lines the photon, and dotted lines the noise field. In the above diagrams each electromagnetic vertex gives a factor proportional to $e$ while each noise vertex gives a factor proportional to $\sqrt\gamma$.}
\end{figure}
When computing the square modulus of the transition amplitude the lowest order contributions are those proportional to $e^2\gamma$. These are obtained by taking the square modulus of the sum of the contributions due to the first two diagrams 1a and 1b, plus the product between the complex conjugate of the contribution due to the single-vertex diagram 2 with each one of the contributions corresponding to the three-vertex diagrams 3a, 3b, 3c plus the complex conjugate of this last contribution.

The first two diagrams are expected from a naive application of the Feynman rules. They are the only ones which have  been considered in~\cite{ref:fu,ref:abd}.  The reason is that, for a free particle ($\omega_0=0$), one can always choose an initial state of zero momentum. In this case, diagram 2 vanishes. Moreover, in such a situation also diagram 1b vanishes. This implies that the only non-zero contribution comes from diagram 1a, as computed in~\cite{ref:fu,ref:abd}.

In order to determine the probability for the particle to emit  one photon with wave vector $\mathbf{k}$ and polarization $\lambda$,
one has to take the square modulus of the transition amplitude $T_{fi}$, average over the noise, and, since one is not interested in the final state of the particle, sum over all its possible final states $f$. Since one is also not interested  in the direction and the polarization of the emitted photon, also a sum over these degrees of freedom has to be taken. This second sum contributes with a factor equal to $8\pi k^{2}$. Finally, in order to infer the emission rate one has to take the time derivative. Putting all pieces together, one has:
\begin{equation}
\frac{d\Gamma}{dk}\;=\;8\pi k^{2}\;\underset{f}{\sum}\;\frac{\partial}{\partial t}\;\mathbb{E}|T_{fi}|^{2}.
\end{equation}

It is now instructive to consider the more general case of non-white noises, since it helps in clarifying the issue at stake. For this one encodes the noise field time correlation by a generic function $f(t-s)$ instead of a Dirac delta function. It has been shown in~\cite{ref:im} that, as far as one keeps the computation at the lowest order in $\gamma$, there are no problems involved in naively extending collapse models to non-white noises simply by generalizing the correlation function of the noise.   

As it was shown in detail in~\cite{ref:abd} and the final result for the emission rate is given by:
\begin{equation}
\frac{d\Gamma}{dk}\;=\;\frac{\lambda\hbar e^{2}}{4\pi^{2}\varepsilon_{0}c^{3}m_{0}^{2}r_{C}^{2}k}\;\left[\tilde{f}(0)+\tilde{f}(\omega_{k})\right],
\label{eq:rate non white}
\end{equation} 
with $\lambda\equiv\gamma/8\pi^{3/2}r_{C}^{3}$ equal to the collapse rate first introduced in the GRW model~\cite{ref:grw}, and: 
\begin{equation}\label{eq:ftilda}
\tilde{f}(k) \equiv \int_{- \infty}^{+ \infty} f(s) e^{iks} ds,
\end{equation}
with $f(s)$ being the noise field time correlation. In the white noise case $f(s)=\delta(s)$, which implies $\tilde{f}(k)=1$. 

The second term of Eq.~\eqref{eq:rate non white} is the expected one: the probability of emitting a photon with momentum $k$ is proportional to the weight of the Fourier component of the noise correlation corresponding to the frequency $\omega_{k}=kc$. On the other hand, the first term is independent of the photon's momentum. It is related to the zero energy component of the field only. This is unexpected as the typical picture is that the noise gives energy to the particle, and such energy is converted into that of the emitted photon. Therefore, it should not be the case that the zero energy component of the field allows to emit a photon with an arbitrarily high energy.

One can object that this is a peculiar feature of the CSL model which differs from the standard quantum picture. This criticism can be easily rejected by remembering that, in order to perform the computation in an easier way, one has effectively considered a standard quantum Hamiltonian with a random term.

\section{Main Result}

As anticipated in the introduction, in~\cite{ref:abd} a first way out to the problem was found. By repeating the calculation with using a wave packet to represent the final state of the particle and with the noise being confined to a bounded region, the undesired term disappears. This means that this term comes from a naive use of perturbation theory. The origin of the unphysical term was investigated in~\cite{ref:ad} for the simple QMUPL model, showing that it arises when higher order terms in the electromagnetic interaction are neglected. Here we prove this result for the CSL model, which is structurally different from the QMUPL model. Technical details are reproduced in the next section. In this section we discuss the main result and conclusions.

We define the unperturbed and perturbed Hamiltonian as follows:
\begin{eqnarray}
{H}_{0} & = & \frac{{\bf p}^2}{2m} + \frac{1}{2}
m\omega_0^2\, {\bf q}^2 + \frac{1}{2}\epsilon_0 \int d^3x \, \left[ {\bf
E}_{\perp}^2 + c^2 {\bf B}^2 \right] -\frac{e}{m}\mathbf{A}\cdot\mathbf{p}+\frac{e^{2}}{2m}\mathbf{A}^{2}  \label{eq:one}
\\
{H}_{1} & = & -\hbar \frac{\sqrt{\gamma}}{m_{0}} \int  d^{3}x \,M({\bf x})\xi_{t}(\mathbf{x}).
\label{eq:two}
\end{eqnarray}
This means that, differently from~\cite{ref:fu,ref:abd}, we will solve the Heisenberg equations of motion exactly (except for the dipole approximation) for the free kinetic term  of the particle and the {\it full} electromagnetic term. In this way we automatically include higher order terms in the electromagnetic field in our analysis, which will prove to be crucial. The noise terms will be analyzed only to first perturbative order. Differently from~\cite{ref:abd}, we will not confine the noise to a bounded region, and our result will be independent of the initial and final state of the particle.

Most integrals can be solved exactly. We identify those terms, which at lowest order in $e$ give a finite contribution to the asymptotic emission rate, while they decay exponentially in time when higher order corrections are included, as in our case. Such terms are precisely those responsible for the appearance of the term $\tilde{f}(0)$ in Eq.~\eqref{eq:rate non white}. When such terms are appropriately taken into account the formula for the emission rate becomes:
\begin{equation}\label{eq:finalharmonic}
\frac{d\Gamma}{dk}=\frac{e^{2}\hbar\lambda c}{4\pi^{2}\epsilon_{0}m_{0}^{2}r_{C}^{2}}\frac{k^{3}}{\left(\omega_{k}^{2}-\omega_{0}^{2}\right)^{2}}\tilde{f}\left(\omega_{k}\right),
\end{equation}
where only terms proportional to $\tilde{f}\left(\omega_{k}\right)$ survive. This formula, in the free particle limit $\omega_{0}\to0$, reduces to:
\begin{equation}\label{eq:finalfree}
\frac{d\Gamma}{dk}=\frac{e^{2}\hbar\lambda}{4\pi^{2}\epsilon_{0}m_{0}^{2}r_{C}^{2}c^{3}k}\tilde{f}\left(\omega_{k}\right),
\end{equation}
which is the expected result. 

This result shows that, as anticipated in the work of~\cite{ref:ad} with the QMUPL model, the suspicious term in Eq.~\eqref{eq:rate non white} arises because calculations were limited to the lowest perturbative order. The picture which emerges is the following. The noise acts always in time, not just for a short period as in standard two body interactions. Therefore, even if it is weak its effect builds up constantly, and in the large time limit (the one always considered in the literature) it makes the perturbative analysis untenable. The analysis of~\cite{ref:abd} proved that by taking final wave packets in place of plane waves,  and by confining the noise to a bounded region, is a way to constraint the effect of the noise---so to say, to make it formally equivalent to a standard scattering process---so that perturbative techniques can be safely applied. Our analysis shows clearly that the true behavior can be understood only by taking higher order terms into account. These terms are dominant in the large time limit and force all undesired terms  to vanish exponentially.

We point out that, in order to derive the correct result, it is necessary to bound the particle with a (harmonic) potential, no matter how weakly. Only at the end of all calculations, the free particle limit can be taken. This procedure is perfectly consistent with physical reality, where free particles which are perfectly free forever do not exist. On the mathematical level however it shows that for a perfectly free particle, even when higher order terms are included in the perturbative analysis, the undesired term remains. Hence, in the large time limit, the zero component of the noise has all the time to induce photon emission at any frequency. This result is consistent with the analysis of~\cite{ref:ad} for the QMUPL model. But again this is only a mathematical truth, which does not apply to the physical world.

\section{Computation of the emission rate}

This section is mainly technical. We show how to derive the emission rate formula given in Eq.~(\ref{eq:finalharmonic}).  The relevant steps are the following.
\begin{itemize}
\item In subsection IV.A we  consider the master equation for the density matrix $\rho(t)$, and we use it to compute the expectation value of a generic observable, to first perturbative order in $\gamma$, while the electromagnetic term is treated exactly (modulo the dipole approximation). We specialize to the case of the expectation value of the photon number operator $a_{\mathbf{k},\mu}^{\dagger}a_{\mathbf{k},\mu}\left(t\right)$. We express its time evolution in terms of two quantities, $C(t,t_1)$ and $D(t,t_1,t_2)$ defined in Eqs.~\eqref{eq:Cdef} and~\eqref{eq:Ddef} respectively.
\item In IV.B we consider the explicit time dependence of those operators, which are relevant for computing $C(t,t_1)$ and $D(t,t_1,t_2)$.
\item In IV.C and IV.D we find the analytic expression for $C(t,t_1)$ and $D(t,t_1,t_2)$.
\item In IV.E we compute the expectation value of the photon's number operator $a_{\mathbf{k},\mu}^{\dagger}a_{\mathbf{k},\mu}\left(t\right)$. 
\item In IV.F perform all remaining time integrals, and show which terms vanish, which rapidly oscillate, and which instead give a finite contribution in the large time limit.
\item In IV.G we put all pieces together and draw the conclusion.
\end{itemize}

\subsection{The formula for the photon's number operator}

In the dipole approximation the vector potential is independent of ${\bf x}$:
\begin{equation}
{\bf A}({\bf x},t)\;=\;\sum_{\lambda=1}^{2}\int d{\bf k}\, \alpha_{k}\left[\vec{\epsilon}_{{\bf k},\lambda}\, a_{{\bf k}}+\vec{\epsilon}_{{\bf k},\lambda}^{*}\, a_{{\bf k}}^{\dagger}\right].
\end{equation}
Here, $\alpha_{k}=\sqrt{\hbar/2\varepsilon_{0}\omega_{k}{(2\pi)}^{3}}$, where $\omega_{k}=kc$, and $a_{{\bf k}}$, $a_{{\bf k}}^{\dagger}$ are respectively the annihilation and creation operators of a photon with momentum $\hbar {\bf k}$; they obey the commutation rules $[a_{{\bf k}},a_{{\bf k'}}^{\dagger}]=\delta({{\bf k}-{\bf k'}})$. Finally,  $\vec{\epsilon}_{{\bf k},1}$($\vec{\epsilon}_{{\bf k},2}$) are the polarization vectors oriented along the ${\bf x}$ (${\bf y}$) axis. 

In~\cite{ref:im} it was shown that, to first order in $\gamma$, the master equation for the colored-noise CSL model is: 
\begin{equation}
\frac{d\rho\left(t\right)}{dt}=-\frac{i}{\hbar}\left[H,\rho\left(t\right)\right]+\gamma\sum_{i,j=1}^{N}\int_{0}^{t}dsD_{ij}\left(t,s\right)\left[A_{i},\left[A_{j}\left(s-t\right),\rho\left(t\right)\right]\right],
\end{equation}
where $D_{ij}\left(t,s\right)$ is the correlation between the noise ``$i$" at time $t$ and the noise ``$j$" at time $s$. In our case, the discrete sum becomes an integral over space, the collapse operators are $M\left(\mathbf{x}\right)$ and
the correlation function becomes: $D_{ij}\left(t,s\right)=\delta_{ij}f\left(t-s\right)\rightarrow\delta\left(\mathbf{x}-\mathbf{x}'\right)f\left(t-s\right)$.
Therefore we have:
\begin{equation}
\frac{d\rho\left(t\right)}{dt}=-\frac{i}{\hbar}\left[H,\rho\left(t\right)\right]-\gamma\left(\frac{m}{m_{0}}\right)^{2}\int d\mathbf{x}\int_{0}^{t}dsf\left(t-s\right)\left[g\left(\mathbf{x}-\hat{\mathbf{q}}\right),\left[g\left(\mathbf{x}-\hat{\mathbf{q}}\left(s-t\right)\right),\rho\left(t\right)\right]\right]\label{eq:ro}
\end{equation}
This equation is given in the Schr\"odinger picture. We switch to the interaction picture:
\begin{equation}
|\psi\left(t\right)\rangle _{I} \; = \; U^{\dagger}\left(t\right)\left|\psi\left(t\right)\right\rangle _{S},\qquad
O_{I}\left(t\right) \; = \; U^{\dagger}\left(t\right)O_{S}\left(t\right)U\left(t\right),
\end{equation}
where $U\left(t\right)$ is the time evolution operator generated by the Hamiltonian $H_0$ defined in  Eq.~(\ref{eq:one}). Eq.~(\ref{eq:ro}) becomes (from now on we will omit the subscript ``$I$"):
\begin{equation}
\frac{d\rho\left(t\right)}{dt}=-\gamma\left(\frac{m}{m_{0}}\right)^{2}\int d\mathbf{x}\int_{0}^{t}dsf\left(t-s\right)\left[g\left(\mathbf{x}-\hat{\mathbf{q}}\left(t\right)\right),\left[g\left(\mathbf{x}-\hat{\mathbf{q}}\left(s\right)\right),\rho\left(t\right)\right]\right].\label{eq:roInt}
\end{equation}
The solution, to first order in $\gamma$, is: 
\begin{equation}
\rho\left(t\right)=\rho\left(0\right)-\gamma\left(\frac{m}{m_{0}}\right)^{2}\int_{0}^{t}dt_{1}\int_{0}^{t_{1}}dt_{2}f\left(t_{1}-t_{2}\right)\int d\mathbf{x}\left[g\left(\mathbf{x}-\hat{\mathbf{q}}\left(t_{1}\right)\right),\left[g\left(\mathbf{x}-\hat{\mathbf{q}}\left(t_{2}\right)\right),\rho\left(0\right)\right]\right].
\end{equation}

In general, to compute the expectation of the observable $O$ at time $t$, one needs to compute:
\begin{equation}
\left\langle O\left(t\right)\right\rangle :=\textrm{Tr}\left[\rho\left(t\right)O\left(t\right)\right].
\end{equation}
Using the relation 
\begin{equation}
\textrm{Tr}\left(\left[A_{1},\left[A_{2},\left[A_{3},\left[...\left[A_{n},\rho\left(t_{i}\right)\right]\right]\right]\right]\right]O\right)=\left\langle \psi\left(t_{i}\right)\left|\left[\left[\left[O,A_{1}\right],A_{2}\right],..,A_{n}\right]\right|\psi\left(t_{i}\right)\right\rangle, 
\end{equation}
one can write:
\begin{eqnarray}
\left\langle O\left(t\right)\right\rangle  & = & \left\langle \psi\left(0\right)\left|O\left(t\right)\right|\psi\left(0\right)\right\rangle -\gamma\left(\frac{m}{m_{0}}\right)^{2}\int_{0}^{t}dt_{1}\int_{0}^{t_{1}}dt_{2}f\left(t_{1}-t_{2}\right)\times\nonumber \\
& \times & \int d\mathbf{x}\left\langle \psi\left(0\right)\left|\left[\left[O\left(t\right),g\left(\mathbf{x}-\hat{\mathbf{q}}\left(t_{1}\right)\right)\right],g\left(\mathbf{x}-\hat{\mathbf{q}}\left(t_{2}\right)\right)\right]\right|\psi\left(0\right)\right\rangle \label{eq:Oaverage} 
\end{eqnarray}
In the case of the emission rate, the operator $O$ whose average we want to compute is the photon's number operator $a_{\mathbf{k},\mu}^{\dagger}a_{\mathbf{k},\mu}$. We will focus only on the second term in Eq.~(\ref{eq:Oaverage}) because we are interested only in the emission rate due to the interaction with the noise field. We then have:
\begin{eqnarray}
\langle a_{\mathbf{k},\mu}^{\dagger}a_{\mathbf{k},\mu}\left(t\right)\rangle_{\text{\tiny noise}} & = & -\gamma\left(\frac{m}{m_{0}}\right)^{2}\int_{0}^{t}dt_{1}\int_{0}^{t_{1}}dt_{2}f\left(t_{1}-t_{2}\right)\times\label{eq:aa^t}
\nonumber \\
 & \times & \int d\mathbf{x}\langle \psi\left(0\right) | [ [a_{\mathbf{k},\mu}^{\dagger}\left(t\right)a_{\mathbf{k},\mu}\left(t\right),g\left(\mathbf{x}-\hat{\mathbf{q}}\left(t_{1}\right)\right)], g\left(\mathbf{x}-\hat{\mathbf{q}}\left(t_{2}\right)\right) ] |\psi\left(0\right)\rangle. 
\end{eqnarray}
The double commutator can be rewritten as the sum of four terms:
\begin{eqnarray}
[ [a_{\mathbf{k},\mu}^{\dagger}\left(t\right)a_{\mathbf{k},\mu}\left(t\right),g\left(\mathbf{x}-\hat{\mathbf{q}}\left(t_{1}\right)\right) ],g\left(\mathbf{x}-\hat{\mathbf{q}}\left(t_{2}\right)\right) ] & = &  a_{\mathbf{k},\mu}^{\dagger}\left(t\right)\left[\left[a_{\mathbf{k},\mu}\left(t\right),g\left(\mathbf{x}-\hat{\mathbf{q}}\left(t_{1}\right)\right)\right],g\left(\mathbf{x}-\hat{\mathbf{q}}\left(t_{2}\right)\right)\right] \nonumber \\
 & + &  [a_{\mathbf{k},\mu}^{\dagger}\left(t\right),g\left(\mathbf{y}-\hat{\mathbf{q}}\left(t_{2}\right)\right) ]\left[a_{\mathbf{k},\mu}\left(t\right),g\left(\mathbf{x}-\hat{\mathbf{q}}\left(t_{1}\right)\right)\right] \nonumber \\
 & + & [a_{\mathbf{k},\mu}^{\dagger}\left(t\right),g\left(\mathbf{y}-\hat{\mathbf{q}}\left(t_{1}\right)\right)]\left[a_{\mathbf{k},\mu}\left(t\right),g\left(\mathbf{x}-\hat{\mathbf{q}}\left(t_{2}\right)\right)\right] \nonumber \\
& + & [ [a_{\mathbf{k},\mu}^{\dagger}\left(t\right),g\left(\mathbf{x}-\hat{\mathbf{q}}\left(t_{1}\right)\right)],g\left(\mathbf{x}-\hat{\mathbf{q}}\left(t_{2}\right)\right)]a_{\mathbf{k},\mu}\left(t\right). \nonumber \\ \label{eq:4terms}
& &
\end{eqnarray}
As we will show in the next subsections, it suffices to compute only two different terms:
\begin{eqnarray}
C\left(t,t_{1}\right) & \equiv & \left[a_{\mathbf{k},\mu}\left(t\right),g\left(\mathbf{x}-\hat{\mathbf{q}}\left(t_{1}\right)\right)\right],\label{eq:Cdef}\\
D\left(t,t_{1},t_{2}\right) & \equiv & \left[\left[a_{\mathbf{k},\mu}\left(t\right),g\left(\mathbf{x}-\hat{\mathbf{q}}\left(t_{1}\right)\right)\right],g\left(\mathbf{x}-\hat{\mathbf{q}}\left(t_{2}\right)\right)\right],\label{eq:Ddef}
\end{eqnarray}
all the other ones being directly connected to them.

\subsection{Time evolution of the relevant operators}\label{sec:time-ev-operators}

In order to compute the commutators defined in Eq.~(\ref{eq:Cdef}) and Eq.~(\ref{eq:Ddef}), we need to know the time evolution of the operators $a_{\mathbf{k},\mu}\left(t\right)$ and $\hat{\mathbf{q}}\left(t\right)$. These are the same as those computed in~\cite{ref:bd}, after having set up a proper renormalization procedure, if one sets $\lambda=0$:
\begin{eqnarray}
a_{\mathbf{k}\mu}\left(t\right) & = & e^{-i\omega_{k}t}a_{\mathbf{k}\mu}+\frac{ie}{\sqrt{\hbar\epsilon_{0}}}\frac{g\left(\mathbf{k}\right)}{\sqrt{2\omega_{k}}}\epsilon_{\mathbf{k}\mu}^{j}\left[G_{1}^{+}\left(k,t\right)p_{j}-\kappa G_{0}^{+}\left(k,t\right)q_{j}\right]\nonumber \\
\nonumber \\
 & + & \frac{ie^{2}}{\epsilon_{0}}\frac{g\left(\mathbf{k}\right)}{\sqrt{2\omega_{k}}}\epsilon_{\mathbf{k}\mu}^{j}\sum_{\mu'}\int d\mathbf{k}'\frac{g\left(\mathbf{k}'\right)}{\sqrt{2\omega_{k'}}}\epsilon_{\mathbf{k}'\mu'}^{j}\left[G_{+}^{+}\left(k,k',t\right)a_{\mathbf{k}'\mu'}+G_{-}^{+}\left(k,k',t\right)a_{\mathbf{k}'\mu'}^{\dagger}\right],\label{eq:a}\\
\nonumber \\
q_{i}\left(t\right) & = & \left[1-\kappa F_{1}\left(t\right)\right]q_{i}+F_{0}\left(t\right)p_{i}\nonumber\\
\nonumber \\
& - & e\sqrt{\frac{\hbar}{\epsilon_{0}}}\sum_{\mu'}\int d\mathbf{k}'\frac{g\left(\mathbf{k}'\right)}{\sqrt{2\omega_{k'}}}\epsilon_{\mathbf{k}'\mu'}^{i}\left[G_{1}^{+}\left(k',t\right)a_{\mathbf{k}'\mu'}+G_{1}^{-}\left(k',t\right)a_{\mathbf{k}'\mu'}^{\dagger}\right].\label{eq:q}
\end{eqnarray}
For completeness we also report explicitly the functions $G_{0}$,
$G_{1}$, $F_{0}$ and $F_{1}$ (already introduced in~\cite{ref:bd}), which we will use in the next sections:
\begin{eqnarray}
F_{0}\left(t\right) & := & \sum_{l=1}^{3}e^{z_{l}t}\left[\frac{z-z_{l}}{H\left(z\right)}\right]_{z=z_{l}}= \label{eq:F0}\\ 
& = & -\frac{e^{z_{1}t}}{\beta\left(z_{1}-z_{2}\right)\left(z_{1}-z_{3}\right)}+\frac{e^{z_{2}t}}{\beta\left(z_{1}-z_{2}\right)\left(z_{2}-z_{3}\right)}-\frac{e^{z_{3}t}}{\beta\left(z_{1}-z_{3}\right)\left(z_{2}-z_{3}\right)},\nonumber\\
& &\nonumber\\
F_{1}\left(t\right) & := & \sum_{l=1}^{3}\frac{e^{z_{l}t}}{z_{l}}\left[\frac{z-z_{l}}{H\left(z\right)}\right]_{z=z_{l}}+\frac{1}{\beta z_{1}z_{2}z_{3}}= \label{eq:F1}\\ 
& = & -\frac{e^{z_{1}t}}{\beta z_{1}\left(z_{1}-z_{2}\right)\left(z_{1}-z_{3}\right)}+\frac{e^{z_{2}t}}{\beta z_{2}\left(z_{1}-z_{2}\right)\left(z_{2}-z_{3}\right)}-\frac{e^{z_{3}t}}{\beta z_{3}\left(z_{1}-z_{3}\right)\left(z_{2}-z_{3}\right)}+\frac{1}{\beta z_{1}z_{2}z_{3}},\nonumber\\
& &\nonumber\\
G_{0}^{\pm}\left(k,t\right) & := &  \sum_{l=1}^{3}\frac{e^{z_{l}t}}{\left(z_{l}\pm i\omega_{k}\right)}\left[\frac{z-z_{l}}{H\left(z\right)}\right]_{z=z_{l}}\mp\frac{e^{\mp i\omega_{k}t}}{H\left(\mp i\omega_{k}\right)}= \label{eq:G0}\\ 
& = & -\frac{e^{z_{1}t}}{\beta\left(z_{1}-z_{2}\right)\left(z_{1}-z_{3}\right)\left(z_{1}\pm i\omega_{k}\right)}+\frac{e^{z_{2}t}}{\beta\left(z_{1}-z_{2}\right)\left(z_{2}-z_{3}\right)\left(z_{2}\pm i\omega_{k}\right)}\nonumber\\
&  & -\frac{e^{z_{3}t}}{\beta\left(z_{1}-z_{3}\right)\left(z_{2}-z_{3}\right)\left(z_{3}\pm i\omega_{k}\right)}\mp\frac{e^{\mp i\omega_{k}t}}{\beta\left(z_{1}\pm i\omega_{k}\right)\left(z_{2}\pm i\omega_{k}\right)\left(z_{3}\pm i\omega_{k}\right)},\nonumber\\
& &\nonumber\\
G_{1}^{\pm}\left(k,t\right) & := &  \sum_{l=1}^{3}\frac{z_{l}e^{z_{l}t}}{\left(z_{l}\pm i\omega_{k}\right)}\left[\frac{z-z_{l}}{H\left(z\right)}\right]_{z=z_{l}}\mp\frac{i\omega_{k}e^{\mp i\omega_{k}t}}{H\left(\mp i\omega_{k}\right)}= \label{eq:G1}\\ 
& = & -\frac{z_{1}e^{z_{1}t}}{\beta\left(z_{1}-z_{2}\right)\left(z_{1}-z_{3}\right)\left(z_{1}\pm i\omega_{k}\right)}+\frac{z_{2}e^{z_{2}t}}{\beta\left(z_{1}-z_{2}\right)\left(z_{2}-z_{3}\right)\left(z_{2}\pm i\omega_{k}\right)}\nonumber\\
&  & -\frac{z_{3}e^{z_{3}t}}{\beta\left(z_{1}-z_{3}\right)\left(z_{2}-z_{3}\right)\left(z_{3}\pm i\omega_{k}\right)}\mp\frac{i\omega_{k}e^{\mp i\omega_{k}t}}{\beta\left(z_{1}\pm i\omega_{k}\right)\left(z_{2}\pm i\omega_{k}\right)\left(z_{3}\pm i\omega_{k}\right)}\nonumber
\end{eqnarray}
where:
\begin{equation}
z_{1}=m/\beta\;\;\;\;\; z_{2,3}=-\frac{\omega_{0}^{2}\beta}{2m}\pm i\omega_{0}\;\;\;\;\;\textrm{and}\;\;\;\;\;\beta = \frac{e^2}{6 \pi \epsilon_0 c^3}.\label{eq:z}
\end{equation}
These functions contain terms that depend on the time in different ways. Some terms contain $\exp[z_1 t]$, which includes the well know runaway behavior due to the renormalization procedure. We pragmatically dismiss them, as is standard practice. There are constant and oscillating terms, which  can potentially  give important contributions to the emission rate. More importantly, there are terms containing $\exp[z_2 t]$ or $\exp[z_3 t]$, which vanish for large times.  These are the crucial terms. Performing the computation to the lowest order in the electromagnetic interaction is equivalent to setting $\beta=0$. In such a case the decaying behavior of the exponential functions containing $z_2$ and $z_3$ is lost and the associated terms give a finite contribution to the emission rate. This contribution turns out to be, in the case of a free particle, proportional to $\tilde{f}(0)$. However, if one performs the computation to higher order their contribution vanishes. This shows that, in order to get the correct result for the radiation emission, it is important not to treat the electromagnetic interaction only at the first order. 

\subsection{Analytic expression of $C\left(t,t_{1}\right)$}

In order to compute the commutators entering Eq.~(\ref{eq:Cdef}) explicitly, it is worthwhile to find some general computational rule. Consider a family of operators $A$, $B$, $C_{1}$, ...,$C_{n}$ such that all the commutators of pairs of these operators vanish apart from $\left[A,B\right]=\epsilon$. Let us define
\begin{equation}
O:=aA+bB+\sum_{j=1}^{n}c_{j}C_{j}
\end{equation}
where $a$, $b$ and $c_{j}$ are constants or functions of the time (the discrete sum could be also an integral on some continuous parameter
as in the case of integrals containing $a_{\mathbf{k}\mu}^{\dagger}$ and $a_{\mathbf{k}\mu}$). Then
\begin{equation}
\left[A,e^{\alpha O^{2}}\right]=\sum_{k=0}^{\infty}\frac{\alpha^{k}}{k!}\left[A,O^{2k}\right]
\end{equation}
and one can show that: 
\begin{equation}
\left[A,O^{2k}\right]=2b\epsilon kO^{2k-1},
\end{equation}
so that the commutator becomes:
\begin{equation}
\left[A,e^{\alpha O^{2}}\right]=\sum_{k=1}^{\infty}\frac{\alpha^{k}}{k!}2b\epsilon kO^{2k-1}=2b\epsilon\alpha O\sum_{k=1}^{\infty}\frac{\alpha^{k-1}}{\left(k-1\right)!}O^{2k-2}=\left[A,O\right]2\alpha Oe^{\alpha O^{2}}.\label{eq:comm}
\end{equation}
Using Eq.~(\ref{eq:comm}) we have:
\begin{equation}
C\left(t,t_{1}\right)  =  \left[a_{\mathbf{k},\mu}\left(t\right),g\left(\mathbf{x}-\hat{\mathbf{q}}\left(t_{1}\right)\right)\right]=\frac{1}{\left(\sqrt{2\pi}r_{C}\right)^{3}}\left[a_{\mathbf{k},\mu}\left(t\right),\prod_{i=1}^{3}\exp\left[-\frac{\left(x_{i}-\hat{q}_{i}\left(t_{1}\right)\right)^{2}}{2r_{C}^{2}}\right]\right]=\label{eq:C}
\end{equation}
\begin{eqnarray}
 & = & \frac{1}{\left(\sqrt{2\pi}r_{C}\right)^{3}}\exp\left[-\frac{\left(x_{1}-\hat{q}_{1}\left(t_{1}\right)\right)^{2}}{2r_{C}^{2}}\right]\exp\left[-\frac{\left(x_{2}-\hat{q}_{2}\left(t_{1}\right)\right)^{2}}{2r_{C}^{2}}\right]\left[a_{\mathbf{k},\mu}\left(t\right),\exp\left[-\frac{\left(x_{3}-\hat{q}_{3}\left(t_{1}\right)\right)^{2}}{2r_{C}^{2}}\right]\right]\nonumber \\
 & &\nonumber\\
 & + & \frac{1}{\left(\sqrt{2\pi}r_{C}\right)^{3}}\exp\left[-\frac{\left(x_{1}-\hat{q}_{1}\left(t_{1}\right)\right)^{2}}{2r_{C}^{2}}\right]\left[a_{\mathbf{k},\mu}\left(t\right),\exp\left[-\frac{\left(x_{2}-\hat{q}_{2}\left(t_{1}\right)\right)^{2}}{2r_{C}^{2}}\right]\right]\exp\left[-\frac{\left(x_{3}-\hat{q}_{3}\left(t_{1}\right)\right)^{2}}{2r_{C}^{2}}\right]\nonumber \\
 & &\nonumber\\
 & + & \frac{1}{\left(\sqrt{2\pi}r_{C}\right)^{3}}\left[a_{\mathbf{k},\mu}\left(t\right),\exp\left[-\frac{\left(x_{1}-\hat{q}_{1}\left(t_{1}\right)\right)^{2}}{2r_{C}^{2}}\right]\right]\exp\left[-\frac{\left(x_{2}-\hat{q}_{2}\left(t_{1}\right)\right)^{2}}{2r_{C}^{2}}\right]\exp\left[-\frac{\left(x_{3}-\hat{q}_{3}\left(t_{1}\right)\right)^{2}}{2r_{C}^{2}}\right].\nonumber 
\end{eqnarray}
This means that we just have to focus on:
\begin{equation}
C_{i}\left(t,t_{1}\right)\equiv\left[a_{\mathbf{k},\mu}\left(t\right),\exp\left[-\frac{\left(x_{i}-\hat{q}_{i}\left(t_{1}\right)\right)^{2}}{2r_{C}^{2}}\right]\right].\label{eq:C_i}
\end{equation}
Using Eq.~(\ref{eq:comm}), it is easy to see that the term
proportional to $e^{2}$ in $a_{\mathbf{k}\mu}\left(t\right)$ (see the second line of Eq.~(\ref{eq:a})), when
commuted with $\exp\left[-\frac{\left(x_{i}-\hat{q}_{i}\left(t_{1}\right)\right)^{2}}{2r_{C}^{2}}\right]$, gives a term of order $e^{3}$. Since in the end we want a result at the lowest order $e^2$, we can neglect this contribution. So we can write $C_{i}\left(t,t_{1}\right)$ as the sum of two terms:
\begin{equation}
C_{i}\left(t,t_{1}\right) = C_{1i}\left(t,t_{1}\right)+C_{2i}\left(t,t_{1}\right).
\end{equation}
Using Eq.~(\ref{eq:comm}) we compute:
\begin{eqnarray}
C_{1i}\left(t,t_{1}\right) & \equiv & e^{-i\omega_{k}t}\left[a_{\mathbf{k}\mu},\exp\left[-\frac{\left(x_{i}-\hat{q}_{i}\left(t_{1}\right)\right)^{2}}{2r_{C}^{2}}\right]\right]=\nonumber \\
\nonumber \\
 & = & -\frac{e\hbar g\left(\mathbf{k}\right)}{\sqrt{\hbar\epsilon_{0}2\omega_{k}}r_{C}^{2}}\epsilon_{\mathbf{k}\mu}^{i}e^{-i\omega_{k}t}G_{1}^{-}\left(k,t_{1}\right)\left(x_{i}-\hat{q}_{i}\left(t_{1}\right)\right)\exp\left[-\frac{\left(x_{i}-\hat{q}_{i}\left(t_{1}\right)\right)^{2}}{2r_{C}^{2}}\right],\\
\nonumber \\
C_{2i}\left(t,t_{1}\right) & \equiv & \frac{ie}{\sqrt{\hbar\epsilon_{0}}}\frac{g\left(\mathbf{k}\right)}{\sqrt{2\omega_{k}}}\epsilon_{\mathbf{k}\mu}^{j}\left[G_{1}^{+}\left(k,t\right)p_{j}-\kappa G_{0}^{+}\left(k,t\right)q_{j},\exp\left[-\frac{\left(x_{i}-\hat{q}_{i}\left(t_{1}\right)\right)^{2}}{2r_{C}^{2}}\right]\right]\\
\nonumber \\
 & = & \frac{e\hbar g\left(\mathbf{k}\right)}{\sqrt{\hbar\epsilon_{0}2\omega_{k}}r_{C}^{2}}\epsilon_{\mathbf{k}\mu}^{i}\left\{ G_{1}^{+}\left(k,t\right)\left[1-\kappa F_{1}\left(t_{1}\right)\right]+\kappa G_{0}^{+}\left(k,t\right)F_{0}\left(t_{1}\right)\right\}\times \nonumber
\nonumber \\
&\times& \left(x_{i}-\hat{q}_{i}\left(t_{1}\right)\right)\exp\left[-\frac{\left(x_{i}-\hat{q}_{i}\left(t_{1}\right)\right)^{2}}{2r_{C}^{2}}\right].
\end{eqnarray}
Hence, we get:
\begin{eqnarray}
C_{i}\left(t,t_{1}\right) & = & \frac{e\hbar g\left(\mathbf{k}\right)}{\sqrt{\hbar\epsilon_{0}2\omega_{k}}r_{C}^{2}}\left\{ -e^{-i\omega_{k}t}G_{1}^{-}\left(k,t_{1}\right)+G_{1}^{+}\left(k,t\right)\left[1-\kappa F_{1}\left(t_{1}\right)\right]+\kappa G_{0}^{+}\left(k,t\right)F_{0}\left(t_{1}\right)\right\}\times \nonumber \\
\nonumber \\
 & \times & \epsilon_{\mathbf{k}\mu}^{i}\left(x_{i}-\hat{q}_{i}\left(t_{1}\right)\right)\exp\left[-\frac{\left(x_{i}-\hat{q}_{i}\left(t_{1}\right)\right)^{2}}{2r_{C}^{2}}\right].
\end{eqnarray}
 Inserting this in Eq.~(\ref{eq:C}) one gets:
\begin{eqnarray}
C\left(t,t_{1}\right) & = & \frac{e\hbar g\left(\mathbf{k}\right)}{\sqrt{\hbar\epsilon_{0}2\omega_{k}}r_{C}^{2}}\left\{ -e^{-i\omega_{k}t}G_{1}^{-}\left(k,t_{1}\right)+G_{1}^{+}\left(k,t\right)\left[1-\kappa F_{1}\left(t_{1}\right)\right]+\kappa G_{0}^{+}\left(k,t\right)F_{0}\left(t_{1}\right)\right\}\times \nonumber \\
 & \times & \epsilon_{\mathbf{k}\mu}^{j}\left(x_{j}-\hat{q}_{j}\left(t_{1}\right)\right)g\left(\mathbf{x}-\hat{\mathbf{q}}\left(t_{1}\right)\right).\label{eq:Cfinal1}
\end{eqnarray}
where we used the fact that $\left[f\left(\hat{q}_{j}\left(t_{1}\right)\right),g\left(\hat{q}_{k}\left(t_{1}\right)\right)\right]=U^{\dagger}\left(t_{1}\right)\left[f\left(\hat{q}_{j}\right),g\left(\hat{q}_{k}\right)\right]U\left(t_{1}\right)=0$.

The commutator $[a_{\mathbf{k},\mu}^{\dagger}\left(t\right),g\left(\mathbf{x}-\hat{\mathbf{q}}\left(t_{1}\right)\right)]$
is related to the one here above. In fact since:
\begin{equation}
g^{\dagger}\left(\mathbf{x}-\hat{\mathbf{q}}\left(t_{1}\right)\right)=g\left(\mathbf{x}-\hat{\mathbf{q}}\left(t_{1}\right)\right)\;\;\;\textrm{and}\;\;\;\left(a_{\mathbf{k},\mu}\left(t\right)\right)^{\dagger}=a_{\mathbf{k},\mu}^{\dagger}\left(t\right)
\end{equation}
it follows that:
\begin{equation}
[a_{\mathbf{k},\mu}^{\dagger}\left(t\right),g\left(\mathbf{x}-\hat{\mathbf{q}}\left(t_{1}\right)\right)] = -\left[a_{\mathbf{k},\mu}\left(t\right),g\left(\mathbf{x}-\hat{\mathbf{q}}\left(t_{1}\right)\right)\right]^{\dagger}=-C^{\dagger}\left(t,t_{1}\right).
\end{equation}

\subsection{Analytic expression of $D\left(t,t_{1},t_{2}\right)$}

We can use the previous result to compute:
\begin{eqnarray}
D\left(t,t_{1},t_{2}\right) & = & \left[\left[a_{\mathbf{k},\mu}\left(t\right),g\left(\mathbf{x}-\hat{\mathbf{q}}\left(t_{1}\right)\right)\right],g\left(\mathbf{x}-\hat{\mathbf{q}}\left(t_{2}\right)\right)\right]=\nonumber \\
 & = & \frac{e\hbar g\left(\mathbf{k}\right)}{\sqrt{\hbar\epsilon_{0}2\omega_{k}}r_{C}^{2}}\left\{ -e^{-i\omega_{k}t}G_{1}^{-}\left(k,t_{1}\right)+G_{1}^{+}\left(k,t\right)\left[1-\kappa F_{1}\left(t_{1}\right)\right]+\kappa G_{0}^{+}\left(k,t\right)F_{0}\left(t_{1}\right)\right\}\times \nonumber \\
 & \times & \epsilon_{\mathbf{k}\mu}^{j}\left[\left(x_{j}-\hat{q}_{j}\left(t_{1}\right)\right)g\left(\mathbf{x}-\hat{\mathbf{q}}\left(t_{1}\right)\right),g\left(\mathbf{x}-\hat{\mathbf{q}}\left(t_{2}\right)\right)\right].\label{eq:doppio comm}
\end{eqnarray}
Because of the complicated time evolution of $\hat{\mathbf{q}}\left(t\right)$ given by Eq.~(\ref{eq:q}), the commutator in Eq.~(\ref{eq:doppio comm}) is hard to evaluate. However, by looking at Eq.~(\ref{eq:aa^t}), we can see that we only need to compute the expectation value of the commutator $D\left(t,t_{1},t_{2}\right)$ on the initial state, and integrate over $\mathbf{x}$,
i.e. we only need to compute:
\begin{equation}
I:=\int d\mathbf{x}\left\langle \psi\left(0\right)\left|\left(x_{j}-\hat{q}_{j}\left(t_{1}\right)\right)g\left(\mathbf{x}-\hat{\mathbf{q}}\left(t_{1}\right)\right)g\left(\mathbf{x}-\hat{\mathbf{q}}\left(t_{2}\right)\right)\right|\psi\left(0\right)\right\rangle. 
\end{equation}
The other term of the commutator in Eq.~(\ref{eq:doppio comm}) involve an integral over $\mathbf{x}$ that is equal to $I^*$.
Let us rewrite $I$ in the following way:
\begin{eqnarray}
I & = & \int d\mathbf{x}\left\langle \psi\left(0\right)\left|\left(x_{j}-\hat{q}_{j}\left(t_{1}\right)\right)g\left(\mathbf{x}-\hat{\mathbf{q}}\left(t_{1}\right)\right)g\left(\mathbf{x}-\hat{\mathbf{q}}\left(t_{2}\right)\right)\right|\psi\left(0\right)\right\rangle =\nonumber \\
 & = & \int d\mathbf{x} \langle \psi\left(0\right) | U^{\dagger}\left(t_{1}\right)\left(x_{j}-\hat{q}_{j}\right)g\left(\mathbf{x}-\hat{\mathbf{q}}\right)U\left(t_{1}\right)U^{\dagger}\left(t_{2}\right)g\left(\mathbf{x}-\hat{\mathbf{q}}\right)U\left(t_{2}\right) | \psi\left(0\right) \rangle =\nonumber \\
 & = & \int d\mathbf{q}\int d\mathbf{q}'\int d\mathbf{x}\left(x_{j}-q_{j}\right)g\left(\mathbf{x}-\mathbf{q}\right)g\left(\mathbf{x}-\mathbf{q}'\right)\times\nonumber \\
 & \times & \langle \psi\left(0\right) | U^{\dagger}\left(t_{1}\right) | \mathbf{q}\rangle \langle \mathbf{q} | U\left(t_{1}\right)U^{\dagger}\left(t_{2}\right) | \mathbf{q}'\rangle \langle \mathbf{q}' | U\left(t_{2}\right) | \psi\left(0\right)\rangle. 
\end{eqnarray}
The integral over $\mathbf{x}$ can be computed introducing $\mathbf{z}:=\mathbf{x}-\frac{\mathbf{q}+\mathbf{q}'}{2}$
and then defining $\mathbf{a}:=\frac{\mathbf{q}-\mathbf{q}'}{2}$
so that $\mathbf{x}-\mathbf{q}=\mathbf{z}-\mathbf{a}$ and $\mathbf{x}-\mathbf{q}'=\mathbf{z}+\mathbf{a}$. Then, 
\begin{eqnarray}
&&\int d\mathbf{x}\left(x_{j}-q_{j}\right)g\left(\mathbf{x}-\mathbf{q}\right)g\left(\mathbf{x}-\mathbf{q}'\right)=\int d\mathbf{z}\left(z_{j}-a_{j}\right)g\left(\mathbf{z}-\mathbf{a}\right)g\left(\mathbf{z}-\mathbf{a}\right)=\nonumber \\
 &&=\frac{1}{\left(\sqrt{2\pi}r_{C}\right)^{6}}\int d\mathbf{z}\left(z_{j}-a_{j}\right)\exp\left\{ -\frac{1}{2r_{C}^{2}}\left[\left(\mathbf{z}-\mathbf{a}\right)^{2}+\left(\mathbf{z}+\mathbf{a}\right)^{2}\right]\right\} =\nonumber \\
 &&=\frac{1}{\left(\sqrt{2\pi}r_{C}\right)^{6}}\exp\left(-\frac{\mathbf{a}^{2}}{r_{C}^{2}}\right)\int d\mathbf{z}\left(z_{j}-a_{j}\right)\exp\left(-\frac{\mathbf{z}^{2}}{r_{C}^{2}}\right)=\nonumber \\
 &&=-\frac{a_{j}}{\left(\sqrt{2\pi}r_{C}\right)^{6}}\exp\left(-\frac{\mathbf{a}^{2}}{r_{C}^{2}}\right)\pi^{3/2}r_{C}^{2}=-\frac{\left(q_{j}-q_{j}'\right)}{2\left(\sqrt{2\pi}r_{C}\right)^{6}}\exp\left(-\frac{\left(\mathbf{q}-\mathbf{q}'\right)^{2}}{4r_{C}^{2}}\right)\pi^{3/2}r_{C}^{2}. 
\end{eqnarray}
This means that:
\begin{eqnarray}
I & = & -\frac{\pi^{3/2}r_{C}^{2}}{2\left(\sqrt{2\pi}r_{C}\right)^{6}}\int d\mathbf{q}\int d\mathbf{q}'\left(q_{j}-q_{j}'\right)\exp\left(-\frac{\left(\mathbf{q}-\mathbf{q}'\right)^{2}}{4r_{C}^{2}}\right)\times\nonumber\\
 & \times & \langle \psi\left(0\right) | U^{\dagger}\left(t_{1}\right) | \mathbf{q}\rangle \langle \mathbf{q} | U\left(t_{1}\right)U^{\dagger}\left(t_{2}\right) | \mathbf{q}'\rangle \langle \mathbf{q}' | U\left(t_{2}\right) | \psi\left(0\right) \rangle \; \simeq \; 0.
\end{eqnarray}
In the last step we used the fact that the Gaussian is relevant only when $\mathbf{q}\simeq\mathbf{q}'$, but in such a case the term $\left(q_{j}-q_{j}'\right)$ becomes small. On the contrary, as we will see in the next section, there are other contributions that are non negligible even when $\mathbf{q}\simeq\mathbf{q}'$. The fact that $I$ is negligible means that the first term in Eq.~(\ref{eq:4terms}), containing this double commutator, can be neglected. The same holds for the fourth term, being it the complex conjugate of the first one.

\subsection{Computation of the average photon number}

Using the results of the previous subsections, we can write the commutators in Eq.~(\ref{eq:4terms}) in the following way way:
\begin{equation}
[ [a_{\mathbf{k},\mu}^{\dagger}\left(t\right)a_{\mathbf{k},\mu}\left(t\right),g\left(\mathbf{x}-\hat{\mathbf{q}}\left(t_{1}\right)\right) ],g\left(\mathbf{x}-\hat{\mathbf{q}}\left(t_{1}\right)\right) ] = -C^{\dagger}\left(t,t_{2}\right)C\left(t,t_{1}\right)-C^{\dagger}\left(t,t_{1}\right)C\left(t,t_{2}\right)
\end{equation}
with $C\left(t,t'\right)$ defined in Eq.~(\ref{eq:Cfinal1}). Accordingly, Eq.~(\ref{eq:aa^t}) becomes:
\begin{eqnarray}
\langle a_{\mathbf{k},\mu}^{\dagger}a_{\mathbf{k},\mu}\left(t\right)\rangle_{\text{\tiny noise}} & = & \gamma\left(\frac{m}{m_{0}}\right)^{2}\int_{0}^{t}dt_{1}\int_{0}^{t_{1}}dt_{2}f\left(t_{1}-t_{2}\right)\times\label{eq:aa^t2} \\
 & \times & \int d\mathbf{x} \; \langle \psi\left(0\right) | C^{\dagger}\left(t,t_{2}\right)C\left(t,t_{1}\right)+C^{\dagger}\left(t,t_{1}\right)C\left(t,t_{2}\right) | \psi\left(0\right) \rangle \nonumber \\
 & = & \gamma\left(\frac{m}{m_{0}}\right)^{2}\int_{0}^{t}dt_{1}\int_{0}^{t}dt_{2}f\left(t_{1}-t_{2}\right)\int d\mathbf{x} \; \langle \psi\left(0\right) | C^{\dagger}\left(t,t_{2}\right)C\left(t,t_{1}\right) | \psi\left(0\right) \rangle \nonumber
\end{eqnarray}
where in the second line we used the fact that we consider a noise field with symmetric correlation function $f\left(t_{1}-t_{2}\right)=f\left(\left|t_{1}-t_{2}\right|\right)$. By substituting Eq.~(\ref{eq:Cfinal1}) in Eq.~(\ref{eq:aa^t2}), one gets:
\begin{eqnarray}
\langle a_{\mathbf{k},\mu}^{\dagger}a_{\mathbf{k},\mu}\left(t\right)\rangle_{\text{\tiny noise}} & = & \gamma\left(\frac{m}{m_{0}}\right)^{2}\frac{e^{2}\hbar\left|g\left(\mathbf{k}\right)\right|^{2}}{\epsilon_{0}2\omega_{k}r_{C}^{4}}\epsilon_{\mathbf{k}\mu}^{i*}\epsilon_{\mathbf{k}\mu}^{j}\int_{0}^{t}dt_{1}\int_{0}^{t}dt_{2}f\left(t_{1}-t_{2}\right)\times\label{eq:aa^t3} \\
 & \times & \left\{ -e^{i\omega_{k}t}G_{1}^{-*}\left(k,t_{2}\right)+G_{1}^{+*}\left(k,t\right)\left[1-\kappa F_{1}^{*}\left(t_{2}\right)\right]+\kappa G_{0}^{+*}\left(k,t\right)F_{0}^{*}\left(t_{2}\right)\right\}\times \nonumber \\
 & \times & \left\{ -e^{-i\omega_{k}t}G_{1}^{-}\left(k,t_{1}\right)+G_{1}^{+}\left(k,t\right)\left[1-\kappa F_{1}\left(t_{1}\right)\right]+\kappa G_{0}^{+}\left(k,t\right)F_{0}\left(t_{1}\right)\right\} I_{ij}\nonumber
\end{eqnarray}
where:
\begin{equation}
I_{ij}:=\int d\mathbf{x}\left\langle \psi\left(0\right)\left|g\left(\mathbf{x}-\hat{\mathbf{q}}\left(t_{1}\right)\right)\left(x_{i}-\hat{q}_{i}\left(t_{1}\right)\right)\left(x_{j}-\hat{q}_{j}\left(t_{2}\right)\right)g\left(\mathbf{x}-\hat{\mathbf{q}}\left(t_{2}\right)\right)\right|\psi\left(0\right)\right\rangle \label{eq:I}.
\end{equation}

As we did before, we can rewrite $I_{ij}$ by inserting the identity $\int d\mathbf{q}\left|\mathbf{q}\right\rangle\left\langle \mathbf{q} \right|=\boldsymbol{1}$:
\begin{eqnarray}
I_{ij} & = & \int d\mathbf{q}\int d\mathbf{q}'\int d\mathbf{x} \; g\left(\mathbf{x}-\mathbf{q}\right)\left(x_{i}-q_{i}\right)\left(x_{j}-q_{j}'\right)g\left(\mathbf{x}-\mathbf{q}'\right)\times\nonumber \\
 & \times & \langle \psi\left(0\right) | U^{\dagger}\left(t_{1}\right) | \mathbf{q} \rangle \langle \mathbf{q} | U\left(t_{1}\right)U^{\dagger}\left(t_{2}\right) | \mathbf{q}' \rangle \langle \mathbf{q}' | U\left(t_{2}\right) | \psi\left(0\right) \rangle \label{eq:I2}.
\end{eqnarray}
 We compute the integral over $\mathbf{x}$ by performing the change of variable: $\mathbf{z}:=\mathbf{x}-\frac{\mathbf{q}+\mathbf{q}'}{2}$
and introducing $\mathbf{a}:=\frac{\mathbf{q}-\mathbf{q}'}{2}$; we get:
\begin{eqnarray}
X_{ij} & := & \int d\mathbf{x}g\left(\mathbf{x}-\mathbf{q}\right)\left(x_{i}-q_{i}\right)\left(x_{j}-q_{j}'\right)g\left(\mathbf{x}-\mathbf{q}'\right)=\int d\mathbf{z}g\left(\mathbf{z}-\mathbf{a}\right)\left(z_{i}-a_{i}\right)\left(z_{j}+a_{j}\right)g\left(\mathbf{z}+\mathbf{a}\right)=\nonumber\\
 &&=\frac{1}{\left(\sqrt{2\pi}r_{C}\right)^{6}}\int d\mathbf{z}\exp\left\{ -\frac{1}{2r_{C}^{2}}\left[\left(\mathbf{z}-\mathbf{a}\right)^{2}+\left(\mathbf{z}+\mathbf{a}\right)^{2}\right]\right\} \left(z_{i}-a_{i}\right)\left(z_{j}+a_{j}\right)=\nonumber \\
 &&=\frac{1}{\left(\sqrt{2\pi}r_{C}\right)^{6}}\exp\left(-\frac{\mathbf{a}^{2}}{r_{C}^{2}}\right)\int d\mathbf{z}\exp\left(-\frac{\mathbf{z}^{2}}{r_{C}^{2}}\right)\left(z_{i}-a_{i}\right)\left(z_{j}+a_{j}\right).\label{eq:X} 
\end{eqnarray}
Let us focus our attention on the quantity:
\begin{equation}
Z_{ij}:=\int d\mathbf{z}\exp\left(-\frac{\mathbf{z}^{2}}{r_{C}^{2}}\right)\left(z_{i}-a_{i}\right)\left(z_{j}+a_{j}\right).
\end{equation}
If $i\neq j$ the only non zero term is: 
\begin{equation}
Z_{ij}=-a_{i}a_{j}\int d\mathbf{z}\exp\left(-\frac{\mathbf{z}^{2}}{r_{C}^{2}}\right)=-a_{i}a_{j}r_{C}^{3}\pi^{3/2}.
\end{equation}
On the contrary, if $i=j$ we have:
\begin{equation}
Z_{ii}=\int d\mathbf{z}\exp\left(-\frac{\mathbf{z}^{2}}{r_{C}^{2}}\right)\left(z_{i}^{2}-a_{i}^{2}\right)=\pi^{3/2}\frac{r_{C}^{5}}{2}-a_{i}^{2}r_{C}^{3}\pi^{3/2}.
\end{equation}
Therefore,
\begin{eqnarray}
X_{ij} & = & \frac{1}{\left(\sqrt{2\pi}r_{C}\right)^{6}}\exp\left(-\frac{\mathbf{a}^{2}}{r_{C}^{2}}\right)\left[\delta_{ij}\pi^{3/2}\frac{r_{C}^{5}}{2}-a_{i}a_{j}r_{C}^{3}\pi^{3/2}\right]\nonumber \\
\nonumber \\
 & = & \frac{1}{8\pi^{3/2}r_{C}^{3}}\exp\left(-\frac{\left(\mathbf{q}-\mathbf{q}'\right)^{2}}{4r_{C}^{2}}\right)\left[\delta_{ij}\frac{r_{C}^{2}}{2}-\left(\frac{q_{i}-q_{i}'}{2}\right)\left(\frac{q_{j}-q_{j}'}{2}\right)\right].
\end{eqnarray}
Coming back to $I_{ij}$, we have:
\begin{eqnarray}
I_{ij} & = & \frac{1}{8\pi^{3/2}r_{C}^{3}}\int d\mathbf{q}\int d\mathbf{q}'\exp\left(-\frac{\left(\mathbf{q}-\mathbf{q}'\right)^{2}}{4r_{C}^{2}}\right)\left[\delta_{ij}\frac{r_{C}^{2}}{2}-\left(\frac{q_{i}-q_{i}'}{2}\right)\left(\frac{q_{j}-q_{j}'}{2}\right)\right]\times\label{eq:Ifinal}\\
 & \times & \langle \psi\left(0\right) | U^{\dagger}\left(t_{1}\right) | \mathbf{q} \rangle \langle \mathbf{q} | U\left(t_{1}\right)U^{\dagger}\left(t_{2}\right) | \mathbf{q}'\rangle \langle \mathbf{q}' | U\left(t_{2}\right) | \psi\left(0\right) \rangle \simeq\nonumber \\
 & \simeq & \frac{\delta_{ij}}{16\pi^{3/2}r_{C}}\int d\mathbf{q}\int d\mathbf{q}' \langle \psi\left(0\right) | U^{\dagger}\left(t_{1}\right) | \mathbf{q} \rangle \langle \mathbf{q} | U\left(t_{1}\right)U^{\dagger}\left(t_{2}\right) | \mathbf{q}' \rangle \langle \mathbf{q}' | U\left(t_{2}\right) | \psi\left(0\right) \rangle =\frac{\delta_{ij}}{16\pi^{3/2}r_{C}}\nonumber 
\end{eqnarray}
where again we used the fact that, because of the Gaussian weight, the only relevant parts of the integrals are those for which $\mathbf{q}\simeq\mathbf{q}'$.

Inserting Eq.~(\ref{eq:Ifinal}) in Eq.~(\ref{eq:aa^t3}), introducing $\lambda\equiv\frac{\gamma}{8\pi^{3/2}r_{C}^{3}}$ and considering the case of a point particle $g\left(\mathbf{k}\right)=1/\sqrt{\left(2\pi\right)^{3}}$, we get:
\begin{equation}
\langle a_{\mathbf{k},\mu}^{\dagger}a_{\mathbf{k},\mu}\left(t\right)\rangle_{\text{\tiny noise}}=\frac{e^{2}\hbar\lambda m^{2}}{32\pi^{3}\epsilon_{0}m_{0}^{2}\omega_{k}r_{C}^{2}}\;T\left(t\right),\label{eq:aa^t4}
\end{equation}
where we have collected all time dependent factors in:
\begin{eqnarray}
T\left(t\right) & := & \int_{0}^{t}dt_{1}\int_{0}^{t}dt_{2}f\left(t_{1}-t_{2}\right)\left\{ -e^{i\omega_{k}t}G_{1}^{-*}\left(k,t_{2}\right)+G_{1}^{+*}\left(k,t\right)\left[1-\kappa F_{1}^{*}\left(t_{2}\right)\right]+\kappa G_{0}^{+*}\left(k,t\right)F_{0}^{*}\left(t_{2}\right)\right\}\times \nonumber \\
 & \times & \left\{ -e^{-i\omega_{k}t}G_{1}^{-}\left(k,t_{1}\right)+G_{1}^{+}\left(k,t\right)\left[1-\kappa F_{1}\left(t_{1}\right)\right]+\kappa G_{0}^{+}\left(k,t\right)F_{0}\left(t_{1}\right)\right\} \label{eq:T}.
\end{eqnarray}
In the next subsection, we compute them explicitly.

\subsection{Time integrals}

The functions $G_{0}$, $G_{1}$, $F_{0}$ and $F_{1}$ were already defined in Eqs.~(\ref{eq:F0}) to~(\ref{eq:G1}). As anticipated, we neglect the runaway terms, i.e., the ones containing $\exp[{{z_1}t}]$. Since we are interested in the long time behavior, we also neglect all those terms that vanish exponentially in time $t$. What remains is:
\begin{eqnarray}
G_{0}^{\pm}\left(k,t\right) & = & \frac{e^{\mp i\omega_{k}t}}{\left(m\pm i\beta\omega_{k}\right)\left(-\frac{\omega_{0}^{2}\beta}{2m}+i\omega_{0}\pm i\omega_{k}\right)\left(-\frac{\omega_{0}^{2}\beta}{2m}-i\omega_{0}\pm i\omega_{k}\right)},\\
\nonumber \\
G_{1}^{\pm}\left(k,t\right) & = & \mp\frac{i\omega_{k}e^{\mp i\omega_{k}t}}{\left(m\pm i\beta\omega_{k}\right)\left(-\frac{\omega_{0}^{2}\beta}{2m}+i\omega_{0}\pm i\omega_{k}\right)\left(-\frac{\omega_{0}^{2}\beta}{2m}-i\omega_{0}\pm i\omega_{k}\right)}=\mp i\omega_{k}G_{0}^{\pm}\left(k,t\right).
\end{eqnarray}
One can safely use this approximation for terms that depend only on $t$, while for terms under time integrals one has to be more careful and use the full expression which was given in subsection IV.B.
This simplifies the expression for $T(t)$ in the following way:
\begin{eqnarray}
T\left(t\right) & \equiv & \int_{0}^{t}dt_{1}\int_{0}^{t}dt_{2}f\left(t_{1}-t_{2}\right)\left\{ -e^{i\omega_{k}t}G_{1}^{-*}\left(k,t_{2}\right)+G_{0}^{+*}\left(k,t\right)\left[i\omega_{k}-i\omega_{k}\kappa F_{1}^{*}\left(t_{2}\right)+\kappa F_{0}^{*}\left(t_{2}\right)\right]\right\}\times \nonumber\\
 & \times & \left\{ -e^{-i\omega_{k}t}G_{1}^{-}\left(k,t_{1}\right)+G_{0}^{+}\left(k,t\right)\left[-i\omega_{k}+i\omega_{k}\kappa F_{1}\left(t_{1}\right)+\kappa F_{0}\left(t_{1}\right)\right]\right\}\nonumber\\
 & = &T_{A}\left(t\right)+T_{B}\left(t\right)+T_{C}\left(t\right)+T_{D}\left(t\right),\label{eq:T1} 
\end{eqnarray}
where these four terms in the last line are given by:
\begin{eqnarray}
T_{A}\left(t\right) & := & \int_{0}^{t}dt_{1}\int_{0}^{t}dt_{2}f\left(t_{1}-t_{2}\right)\left\{ -e^{i\omega_{k}t}G_{1}^{-*}\left(k,t_{2}\right)\right\} \left\{ -e^{-i\omega_{k}t}G_{1}^{-}\left(k,t_{1}\right)\right\}, \label{eq:T_A}\\
\nonumber \\
T_{B}\left(t\right) & := & \int_{0}^{t}dt_{1}\int_{0}^{t}dt_{2}f\left(t_{1}-t_{2}\right)\left\{ -e^{i\omega_{k}t}G_{1}^{-*}\left(k,t_{2}\right)\right\}\times \nonumber\\
&  & \;\;\;\;\;\;\;\;\;\;\;\;\;\;\;\;\;\;\;\;\;\;\;\;\;\;\;\;\;\;\;\;\;\;\;\,\times\left\{ G_{0}^{+}\left(k,t\right)\left[-i\omega_{k}+i\omega_{k}\kappa F_{1}\left(t_{1}\right)+\kappa F_{0}\left(t_{1}\right)\right]\right\}, \label{eq:T_B}\\
\nonumber \\
T_{C}\left(t\right) & := & \int_{0}^{t}dt_{1}\int_{0}^{t}dt_{2}f\left(t_{1}-t_{2}\right)\left\{ -e^{-i\omega_{k}t}G_{1}^{-}\left(k,t_{1}\right)\right\}\times\nonumber\\
&  &\;\;\;\;\;\;\;\;\;\;\;\;\;\;\;\;\;\;\;\;\;\;\;\;\;\;\;\;\;\;\;\;\;\;\;\,\times\left\{ G_{0}^{+*}\left(k,t\right)\left[i\omega_{k}-i\omega_{k}\kappa F_{1}^{*}\left(t_{2}\right)+\kappa F_{0}^{*}\left(t_{2}\right)\right]\right\},\label{eq:T_C}\\
\nonumber \\
T_{D}\left(t\right) & := & \int_{0}^{t}dt_{1}\int_{0}^{t}dt_{2}f\left(t_{1}-t_{2}\right)\left\{ G_{0}^{+*}\left(k,t\right)\left[i\omega_{k}-i\omega_{k}\kappa F_{1}^{*}\left(t_{2}\right)+\kappa F_{0}^{*}\left(t_{2}\right)\right]\right\}\times \nonumber \\
 &  &\;\;\;\;\;\;\;\;\;\;\;\;\;\;\;\;\;\;\;\;\;\;\;\;\;\;\;\;\;\;\;\;\;\;\;\,\times \left\{ G_{0}^{+}\left(k,t\right)\left[-i\omega_{k}+i\omega_{k}\kappa F_{1}\left(t_{1}\right)+\kappa F_{0}\left(t_{1}\right)\right]\right\}. \label{eq:T_D}
\end{eqnarray}

\noindent 1. Expression for $T_{A}\left(t\right)$. In order to compute the first of the integrals here above, we start by plugging in the explicit expression of $G_{1}^{-}\left(k,t\right)$ given by Eq.~(\ref{eq:G1}). In general, we will encounter integrals having  the following structure:
\begin{eqnarray}
I\left(a,b\right) & \equiv\ & \int_{0}^{t}dt_{1}\int_{0}^{t}dt_{2}f\left(t_{1}-t_{2}\right)e^{at_{1}}e^{bt_{2}}\nonumber \\
 & = & 4\frac{e^{\frac{1}{2}\left(a+b\right)t}}{\left(a+b\right)}\int_{0}^{t}dxf\left(x\right)\sinh\left[\frac{1}{2}\left(a+b\right)\left(t-x\right)\right]\cosh\left[\frac{1}{2}\left(a-b\right)x\right]=\nonumber \\
 & = & 2\frac{e^{\left(a+b\right)t}}{\left(a+b\right)}\int_{0}^{t}dxf\left(x\right)e^{-\frac{1}{2}\left(a+b\right)x}\cosh\left[\frac{1}{2}\left(a-b\right)x\right]\nonumber \\
& - & 2\frac{1}{\left(a+b\right)}\int_{0}^{t}dxf\left(x\right)e^{\frac{1}{2}\left(a+b\right)x}\cosh\left[\frac{1}{2}\left(a-b\right)x\right].\label{eq:I(a,b)}
\end{eqnarray}
In the case of $T_{A}\left(t\right)$, by neglecting again the runaway terms, we have only the following possibilities: $a,b=-\frac{\omega_{0}^{2}\beta}{2m}\pm i\omega_{0}$ or $\pm i\omega_{k}$. If $a+b\neq0$ the integral $I\left(a,b\right)$ either oscillates or is constant. In both cases, it does not contribute to the emission rate. On the contrary, when $a+b=0$ the integral becomes:
\begin{equation}
I\left(a,-a\right)=2\int_{0}^{t}dxf\left(x\right)\left(t-x\right)\cosh\left(ax\right)
\end{equation}
and, as we will see, it contributes to the emission rate. By looking at Eq.~(\ref{eq:T_A}) and Eq.~(\ref{eq:G1}), one can easily see that the only non zero contribution comes from the product of the terms containing $e^{i\omega_{k}t}$ and $e^{-i\omega_{k}t}$. Therefore we get:
\begin{equation}
T_{A}\left(t\right)=\frac{2\omega_{k}^{2}}{\left(m^{2}+\beta^{2}\omega_{k}^{2}\right)\left[\left(\frac{\omega_{0}^{4}\beta^{2}}{4m^{2}}+\omega_{k}^{2}-\omega_{0}^{2}\right)^{2}+\frac{\omega_{0}^{6}\beta^{2}}{m^{2}}\right]}\int_{0}^{t}dxf\left(x\right)\left(t-x\right)\cos\left(\omega_{k}x\right).
\end{equation}
The lowest order contribution is obtained by taking the limit $\beta\rightarrow0$:
\begin{equation}
T_{A}\simeq\frac{2\omega_{k}^{2}}{m^{2}\left(\omega_{k}^{2}-\omega_{0}^{2}\right)^{2}}\int_{0}^{t}dxf\left(x\right)\left(t-x\right)\cos\left(\omega_{k}x\right).\label{eq:T_Afinal}
\end{equation}

\noindent 2. Expression for $T_{B}$$\left(t\right)$ and $T_{C}\left(t\right)$. Let us start with  $T_{B}\left(t\right)$ defined in Eq.~(\ref{eq:T_B}).
Taking out of the integrals the functions that do not depend on 
$t_{1}$ and $t_{2}$ one gets: 
\begin{equation}
T_{B}\left(t\right)=-e^{i\omega_{k}t}G_{0}^{+}\left(k,t\right)\int_{0}^{t}dt_{1}\int_{0}^{t}dt_{2}f\left(t_{1}-t_{2}\right)G_{1}^{-*}\left(k,t_{2}\right)\left[-i\omega_{k}+i\omega_{k}\kappa F_{1}\left(t_{1}\right)+\kappa F_{0}\left(t_{1}\right)\right].\label{eq:T_B2}
\end{equation}
The integrals contain the functions $F_{0}\left(t\right)$, $F_{1}\left(t\right)$ and $G_{1}^{-}\left(k,t\right)$ defined respectively in Eq.~(\ref{eq:F0}), Eq.~(\ref{eq:F1}) and Eq.~(\ref{eq:G1}). As before we will neglect the runaway terms proportional to $e^{z_{1}t}$.
We are left only with integrals having the same structure as that of $I\left(a,b\right)$ defined in Eq.~(\ref{eq:I(a,b)}). Every term which involves $e^{z_{2}t}$ or $e^{z_{3}t}$ does not contribute to the asymptotic rate. This means that when we substitute the functions $F_{0}\left(t\right)$, $F_{1}\left(t\right)$ and $G_{1}^{-}\left(k,t\right)$ in Eq.~(\ref{eq:T_B2}) we can set:
\begin{equation}
F_{0}\left(t_{1}\right)=0\;,\;\;\; F_{1}\left(t_{1}\right)=\frac{1}{\beta z_{1}z_{2}z_{3}}\;\;\;\textrm{and}\;\;\; G_{1}^{-*}\left(k,t_{2}\right)=\left(\frac{i\omega_{k}e^{i\omega_{k}t_{2}}}{\beta\left(z_{1}-i\omega_{k}\right)\left(z_{2}-i\omega_{k}\right)\left(z_{3}-i\omega_{k}\right)}\right)^{*}.
\end{equation}
So we have:
\begin{eqnarray}
T_{B}\left(t\right) & = & -e^{i\omega_{k}t}G_{0}^{+}\left(k,t\right)\left[\frac{\kappa}{\beta z_{1}z_{2}z_{3}}-1\right]\frac{\omega_{k}^{2}}{\beta\left(z_{1}+i\omega_{k}\right)\left(z_{2}+i\omega_{k}\right)\left(z_{3}+i\omega_{k}\right)}\times\nonumber\\
& \times &\int_{0}^{t}dt_{1}\int_{0}^{t}dt_{2}f\left(t_{1}-t_{2}\right)e^{-i\omega_{k}t_{2}}.
\end{eqnarray}
This term gives no contribution at the lowest order.
In fact, if we replace the values of $z_{1}$, $z_{2}$ and $z_{3}$ in the above expression, the contribution of the term in the square brackets is:
\begin{equation}
\left[\frac{\kappa}{\beta z_{1}z_{2}z_{3}}-1\right]=\frac{m\omega_{0}^{2}}{m\left(\frac{\omega_{0}^{4}\beta^{2}}{4m^{2}}+\omega_{0}^{2}\right)}-1=-\frac{\frac{\omega_{0}^{4}\beta^{2}}{4m^{2}}}{\frac{\omega_{0}^{4}\beta^{2}}{4m^{2}}+\omega_{0}^{2}}.\label{eq:[]}
\end{equation}
So by taking again the limit $\beta\rightarrow0$ this term vanishes (the function $G_{0}^{+}\left(k,t\right)$ and the fraction before the integrals give a finite contribution in this limit). Since $T_{C}\left(t\right)=T_{B}^{*}\left(t\right)$, also $T_{C}=0$
at the lowest order.

\noindent 3. Expression for $T_{D}\left(t\right)$. Let us rewrite Eq.~(\ref{eq:T_D}) by taking out of the double integral the functions that do not depend by $t_{1}$ and $t_{2}$:
\begin{eqnarray}
T_{D}\left(t\right)& = &\left|G_{0}^{+}\left(k,t\right)\right|^{2}\int_{0}^{t}dt_{1}\int_{0}^{t}dt_{2}f\left(t_{1}-t_{2}\right)\left[i\omega_{k}-i\omega_{k}\kappa F_{1}^{*}\left(t_{2}\right)+\kappa F_{0}^{*}\left(t_{2}\right)\right]\times\nonumber\\
& \times &\left[-i\omega_{k}+i\omega_{k}\kappa F_{1}\left(t_{1}\right)+\kappa F_{0}\left(t_{1}\right)\right].\label{eq:T_D1}
\end{eqnarray}
The functions $F_{0}\left(t\right)$ and $F_{1}\left(t\right)$ are defined in Eq.~(\ref{eq:F0}) and Eq.~(\ref{eq:F1}). Once again we have only integrals with the same structure as that of $I\left(a,b\right)$ in Eq.~(\ref{eq:I(a,b)}) and therefore the only terms that survive are the ones for which $F_{0}\left(t\right)=0$
and $F_{1}\left(t\right)=\frac{1}{\beta z_{1}z_{2}z_{3}}$. In such a case the above expression becomes:
\begin{equation}
T_{D}\left(t\right)=\left|G_{0}^{+}\left(k,t\right)\right|^{2}\omega_{k}^{2}\left|1-\frac{\kappa}{\beta z_{1}z_{2}z_{3}}\right|^{2}\int_{0}^{t}dt_{1}\int_{0}^{t}dt_{2}f\left(t_{1}-t_{2}\right).\label{eq:T_D1-1}
\end{equation}
This term vanishes in the limit $\beta\rightarrow0$. In fact the term inside the square modulus is proportional to $\beta^{2}$ (see Eq.~(\ref{eq:[]})) while all the others terms remain finite in such a limit.

\subsection{Final Result}

Summarizing, we have computed that at the lowest order $T\left(t\right)=T_{A}\left(t\right)$
with $T_{A}\left(t\right)$ given by Eq.~(\ref{eq:T_Afinal}). Thus, Eq.~(\ref{eq:aa^t4}) simplifies as follows: 
\begin{equation}
\langle a_{\mathbf{k},\mu}^{\dagger}a_{\mathbf{k},\mu}\left(t\right)\rangle_{\text{\tiny noise}}=\frac{e^{2}\hbar\lambda}{32\pi^{3}\epsilon_{0}m_{0}^{2}r_{C}^{2}}\frac{2\omega_{k}}{\left(\omega_{k}^{2}-\omega_{0}^{2}\right)^{2}}\int_{0}^{t}dxf\left(x\right)\left(t-x\right)\cos\left(\omega_{k}x\right).\label{eq:aa^t5}
\end{equation}
The emission rate is obtained by taking the time derivative of the above expression.
Moreover, since $\langle a_{\mathbf{k},\mu}^{\dagger}a_{\mathbf{k},\mu}\left(t\right)\rangle_{\text{\tiny noise}}$
does not depend on the direction and the polarization of the photon
we can sum over these degrees of freedom multiplying by a factor $8\pi k^{2}$. Using 
\begin{equation}
\frac{d}{dt}\int_{0}^{t}dxf\left(x\right)\left(t-x\right)\cos\left(\omega_{k}x\right)=\frac{1}{2}\tilde{f}\left(\omega_{k}\right),
\end{equation}
where the function $\tilde{f}\left(\omega_{k}\right)$ is the one
defined in Eq.~(\ref{eq:ftilda}) we get
\begin{equation}
\frac{d\Gamma}{dk}=\frac{e^{2}\hbar\lambda c}{4\pi^{2}\epsilon_{0}m_{0}^{2}r_{C}^{2}}\frac{k^{3}}{\left(\omega_{k}^{2}-\omega_{0}^{2}\right)^{2}}\tilde{f}\left(\omega_{k}\right).
\end{equation}
This is the desired result.

\section*{Acknowledgements}
A.B. and S.D. acknowledge support from NANOQUESTFIT, the COST Action MP1006 and INFN, Italy. Both of them wish express their deep gratitude to S.L. Adler for many enjoyable and very stimulating discussions on this topic. They wish to acknowledge the hospitality of the Institute for Advanced Study in Princeton and the University of California Davis, respectively, where part of this work was done.

\end{document}